\documentclass[runningheads]{llncs}
\usepackage[T1]{fontenc}
\usepackage{graphicx}
\usepackage[dvipsnames]{xcolor}
\definecolor{Grey}{rgb}{0.5, 0.5, 0.5}
\usepackage[angle=0,pos={0.5\paperwidth,3\topmargin},hanchor=c,vanchor=m]{draftwatermark}
\SetWatermarkText{Preprint of publication at\\International Conference on Web Engineering 2025 (ICWE 2025)}
\SetWatermarkScale{0.6}
\usepackage{booktabs}
\usepackage{listings}
\usepackage{multirow}
\usepackage{caption}
\usepackage{subcaption}
\usepackage{etoolbox}
\expandafter\def\csname ver@subfig.sty\endcsname{}

\PassOptionsToPackage{svgnames,x11names,dvipsnames}{xcolor}
\usepackage[most]{tcolorbox}

\usepackage{alltt}
\usepackage{xspace}
\usepackage{paralist}
\usepackage{comment}
\usepackage{subfig}
\usepackage{amsfonts}
\usepackage{pdfpages}
\usepackage{color,soul}
\usepackage{hyperref}
\usepackage{mathtools}
\usepackage{fp}
\usepackage[utf8]{inputenc}
\usepackage{makecell}
\usepackage{boldline}

\usepackage{listings}
\usepackage{xcolor}

\definecolor{dkgreen}{rgb}{0.0,0.6,0.0}
\definecolor{dkblue}{rgb}{0.0,0.0,0.6}

\lstdefinelanguage{SPARQL}{
  morekeywords={
    SELECT, WHERE, ORDER, BY, DESC, LIMIT, AS
  },
  sensitive=true,
  morecomment=[l]{\#},
}

\newcommand{\code}[1]{\texttt{#1}}

\usepackage[misc,geometry]{ifsym}

\usepackage{enumitem}
\setlist{nosep}

\usepackage[nocompress]{cite}

\newcommand{\ie}{i.e.,\xspace}
\renewcommand{\st}{s.t.,\xspace} 
\newcommand{\eg}{e.g.,\xspace}

\newcommand{\cf}{cf.\xspace}

\newcommand{\qq}[1]{``#1''}

\newcommand{\QALD}{QALD\xspace}
\newcommand{\QALDNinePlus}{QALD-9-plus\xspace}

\newcommand{\Fscore}{F1 score\xspace}

\newcommand{\QA}{QA\xspace}
\newcommand{\NL}{NL\xspace}

\newcommand{\LLM}{LLM\xspace}
\newcommand{\LLMs}{LLMs\xspace}

\newcommand{\KGQA}{KGQA\xspace}
\newcommand{\KG}{KG\xspace}
\newcommand{\KGs}{KGs\xspace}
\newcommand{\SPARQL}{SPARQL\xspace}

\newcommand{\LCQuAD}{LC-QuAD 2.0\xspace}

\newcommand{\Wikidata}{Wikidata\xspace}

\newcommand{\Llama}{Llama\xspace}

\usepackage{todonotes}

\begin{document}
\title{
SPARQL Query Generation with LLMs: Measuring the Impact of Training Data Memorization and Knowledge Injection
}
\titlerunning{SPARQL Query Generation with LLMs}
%
\author{Aleksandr Gashkov\inst{1}\orcidID{0000-0001-6894-2094} \and
Aleksandr Perevalov\inst{1} \orcidID{0000-0002-0877-7063} \and
Maria Eltsova\inst{1,2}\orcidID{0000-0003-3792-8518} \and
Andreas Both\inst{1}\orcidID{0000-0001-7116-9338}
}

\institute{
\includegraphics[height=2ex]{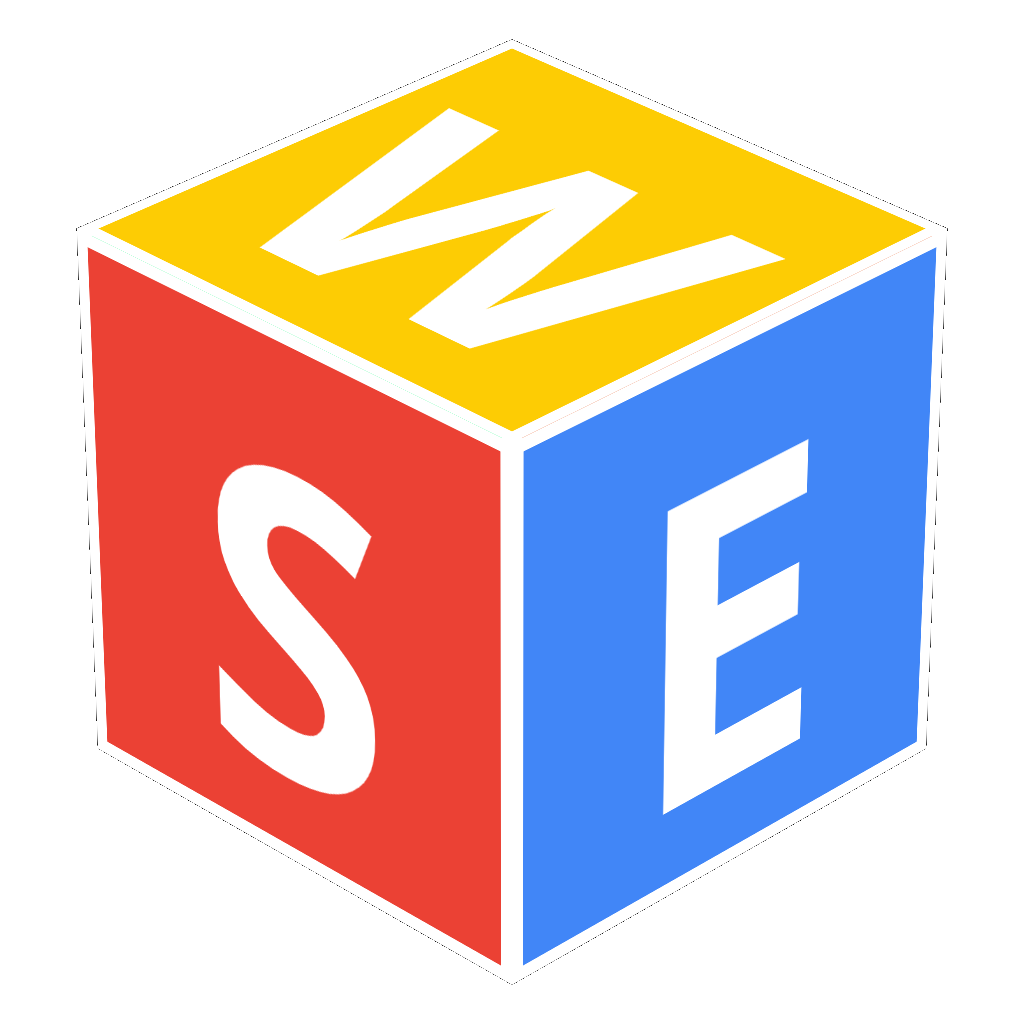} Web \& Software Engineering (WSE) Research Group,\linebreak[1]Leipzig University of Applied Sciences (HTWK Leipzig),
Leipzig, Germany 
\and 
CBZ München GmbH, Heilbronn, Germany 
}
%
\authorrunning{Aleksandr Gashkov, Aleksandr Perevalov, Maria Eltsova, and Andreas Both}
%
%
\maketitle              
\begin{abstract}
Nowadays, the importance of software with natural-language user interfaces cannot be underestimated. 
In particular, in Question Answering (QA) systems, generating a \SPARQL query for a given natural-language question (often named Query Building) from the information retrieved from the same question is the central task of QA systems working over Knowledge Graphs (\KGQA).
Due to the rise of Large Language Models (LLMs), they are considered a well-suited method to increase the quality of the question-answering functionality, as there is still a lot of room for improvement, aiming for enhanced quality and trustworthiness. 
However, LLMs are trained on web data, where researchers have no control over whether the benchmark or the knowledge graph was already included in the training data.
In this paper, we introduce a novel method 
that evaluates the quality of \LLMs by generating a \SPARQL query from a natural-language question under various conditions: (1) zero-shot \SPARQL generation, (2) with knowledge injection, and (3) with \qq{anonymized} knowledge injection.
This enables us, for the first time, to estimate the influence of the training data on the QA quality improved by LLMs.
Ultimately, this will help to identify how portable a method is or whether good results might mostly be achieved because a benchmark was already included in the training data (cf. LLM memorization).
The developed method is portable, robust, and supports any knowledge graph; therefore, it could be easily applied to any \KGQA or LLM, \st generating consistent insights into the actual LLM capabilities is possible. 

\keywords{Question Answering  \and SPARQL query generation \and Large Language Models \and Knowledge Graph \and LLM memorization.}
\end{abstract}

\section{Introduction}\label{sec:introduction}
\newtcbox{\inlinebox}[1][]{enhanced,
 box align=base,
 nobeforeafter,
 colback=white,
 colframe=Grey,
 size=small,
 left=0pt,
 right=0pt,
 boxsep=2pt,
 #1}
\newcommand{\ResearchQuestion}[1]{\inlinebox{RQ#1}\xspace}
\newcommand{\ResearchQuestionA}{\ResearchQuestion{1}}
\newcommand{\ResearchQuestionB}{\ResearchQuestion{2}}
\begin{figure}[t!]
    \centering
    \includegraphics[width=0.9\linewidth]{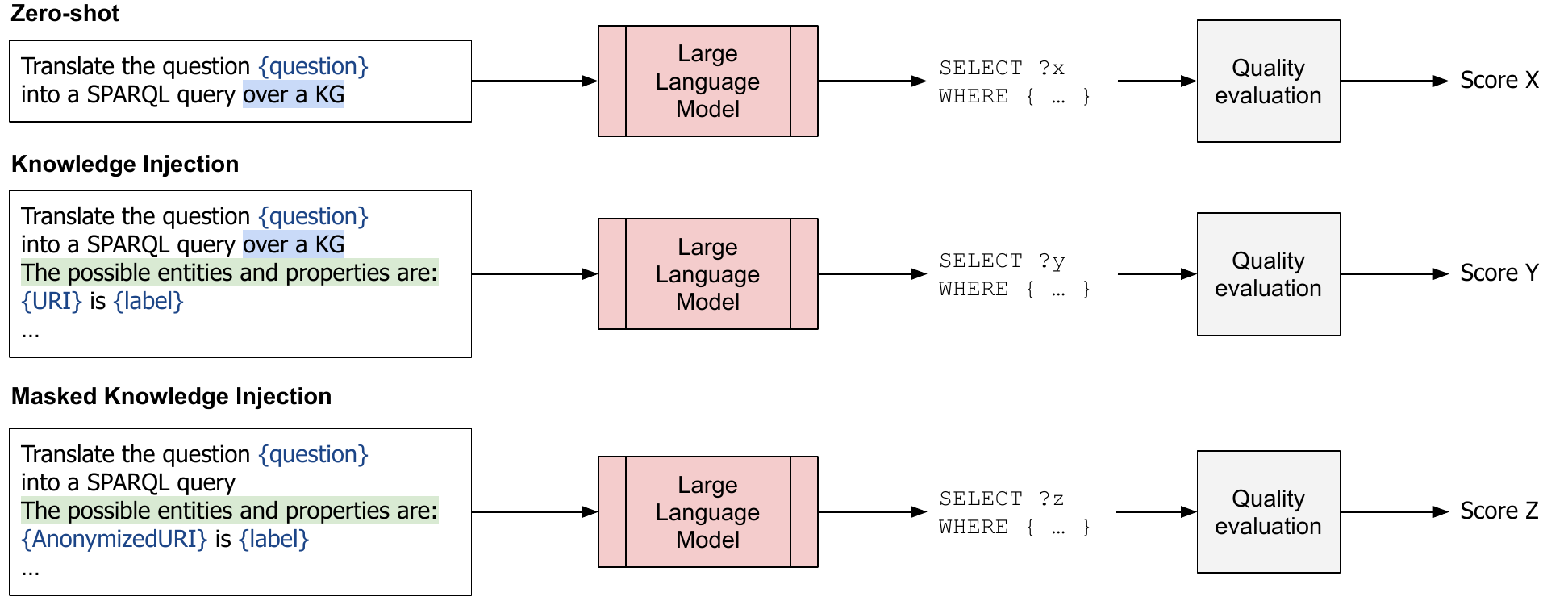}
    \vspace*{-1ex}
    \caption{Overview of the proposed evaluation approach and prompting strategies}
    \label{fig:big-picture}
    \vspace*{-3ex}
\end{figure}

Question Answering (\QA) is aimed at providing a user with precise answers to questions formulated in a natural language (\NL).
\KGQA systems are filling up the gap between Linked Data and end-users by transforming \NL questions into structured queries (\eg represented as \SPARQL\footnote{\href{https://www.w3.org/TR/rdf-sparql-query/}{https://www.w3.org/TR/rdf-sparql-query/}}) to make the information accessible using \NL requests.
However, translating \NL queries into \SPARQL is still a challenge in the field of \KGQA (\cf \cite{diallo2024comprehensive,yani2021challenges,perevalov2024language}).
This task is supposed to be solved with the advent of Large Language Models (\LLMs) that have significantly furthered the field of natural language processing (NLP) lately (\eg~\cite{qin2024large,zhuang2024toolqa,perevalov2024understanding}). 
The rapidly growing number of research papers over the past two years confirms the fact that the possibility of \LLMs to efficiently generate machine-readable queries from the questions written in \NL is being actively discussed.
However, according to \cite{shen2025reasoning}, \LLMs often perform well on entities and relationships with high frequencies but face challenges in less popular topics (long-tail knowledge problem).
Furthermore, the capabilities of LLMs could be questioned as they show anomalies in terms of \emph{memorization} concerning the actual ability to generate correct SPARQL queries obeying a defined context. 
From this context, we derive the following \emph{research questions}:
\begin{description}
    \item[\ResearchQuestionA] Assuming a perfect knowledge injection (\ie in-prompt context definition), how well does the SPARQL query generation perform?
    \item[\ResearchQuestionB] What is the impact of memorization on the SPARQL query generation capabilities of LLMs?
\end{description}
In summary, these research questions are intended to shed light on the capabilities of LLMs in generating SPARQL queries so that software engineers can better assess the usefulness of LLMs for non-popular or private knowledge graphs.

In this paper, we present a novel method for evaluating the quality of \LLMs by generating \SPARQL queries from a \NL question under various conditions (see Fig.~\ref{fig:big-picture} for the general idea): 
    (1) \SPARQL generation without additional information provided to the \LLM (\emph{zero-shot}),
    (2) Providing specific and complete knowledge about the information (\emph{knowledge injection}) required to generate the correct SPARQL query (\eg URI to label mappings), and 
    (3) Reusing knowledge injection with anonymized URIs  (\emph{masked knowledge injection}).
 
In contrast to prior contributions, our work focuses on estimating the influence of the training data on the \QA quality improved by \LLMs.
Our method is ultimately aimed at recognizing of the fact whether good results are only achieved because an \LLM or benchmark was already included in the training data (which might lead to \LLM memorization effects).
In our investigation, we assessed our method using the different (regarding the question structures) \KGQA datasets, \QALDNinePlus~\cite{Perevalov22QALD9plus} and MCWQ~\cite{cui2022compositional}.
In addition, we conducted our experiments on various \LLMs of different sizes, ranging from 7B to 123B parameters.

This paper is structured as follows. 
In Section~\ref{sec:RelatedWork}, we summarize related work on the integration of \LLMs with \KGs for improving semantic parsing and generation of \SPARQL queries from a \NL questions.
Thereafter, in Section~\ref{sec:Mat_and_Met}, we present an overview of our approach, the used datasets, and \LLMs.
Section~\ref{sec:evaluation} evaluates and analyzes the experimental results and erroneous queries produced by \LLMs during the experiments.
The obtained results and limitations are discussed in Section~\ref{sec:limitations}.
Finally, we conclude our work in Section~\ref{sec:conclusion}.
\\[1ex]
\emph{Reproducibility statement}: The source code and data for experiments are available on GitHub\footnote{\url{https://github.com/WSE-research/LLM-generated-SPARQL/}}.

\section{Related Work}\label{sec:RelatedWork}
The integration of \LLMs with various data sources (in particular, knowledge bases, \KGs, and the Web of Data) for solving different \QA tasks in the last two years has evidenced the high scientific interest in this topic in the research community.
Some of the recent approaches try to exploit \LLMs as semantic parsers~(\eg \cite{faria2023question,liu2024proficient, mecharnia2025performance,meyer2024assessing}) directly or by fine-tuning the models.
Mecharnia and d’Aquin~\cite{mecharnia2025performance} presented experiments in fine-tuning \LLMs (\Llama-3-8B, \Llama-2-7B, \Llama-3-70B, and Mixtral-8x7B) for the task of NLQ-to-SPARQL transformation on \QALDNinePlus~\cite{Perevalov22QALD9plus} and QALD-10~\cite{usbeck2024qald}.
Their approach demonstrated promising results (for QALD, the average Macro F1 is near 60\% for all \LLMs).

Meyer et al.~\cite{meyer2024assessing} presented a set of automated benchmarking tasks to assess the basic capabilities of \LLMs to deal with \SPARQL \code{SELECT} queries without special architectures or fine-tuning.
Completing experiments on different benchmarks like well-known \LCQuAD~\cite{dubey2019lc}, CoyPu-Mini~\cite{brei2024leveraging}, Bestiary~\cite{kovriguina2023sparqlgen} and state-of-the-art \LLMs families (Claude, Gemini and GPT), the authors conclude that most evaluated \LLMs have no significant challenges when perceiving \SPARQL \code{SELECT} query syntax or its semantics, however, generating the \code{SELECT} queries with correct semantics still seems to be a difficult task for the models.

Recent research also tried to integrate \LLMs with \KG and other solutions (\eg~\cite{bhandiwad5009450bridging,shen2025reasoning,kovriguina2023sparqlgen,zahera2024generating}).
Zhang et al.~\cite{zhang2025rule} introduced Rule-KBQA, a framework that employs learned rules to guide the generation of logical forms.
Initially, rules are extracted from existing data, after that the authors employed the Rule-Following Fine-Tuned (RFFT) \LLM to generate additional rules, ultimately constructing a comprehensive rule library.
Shen et al.~\cite{shen2025reasoning} suggested a novel framework -- Reasoning with Trees --  that reformulates \KGQA as a discrete decision-making problem, leveraging Monte Carlo Tree Search to iteratively refine reasoning paths.

Zahera et al.~\cite{zahera2024generating} developed an approach to generate \SPARQL queries that employs Chain-of-thoughts (CoT) prompting~\cite{wei2022chainofthoughts} and incorporates entities and relations from the input question.
Their approach evaluated on the \LCQuAD, VQuAnDa~\cite{VQuAnDa2020}, QALD-9, and QALD-10 datasets demonstrates an improvement in \Fscore for both \QALD benchmarks.

Another decision is proposed by Kovriguina et al.~\cite{kovriguina2023sparqlgen} who developed SPARQLGEN -- a one-shot approach for generating \SPARQL queries with prompting \LLMs.
Their approach reached outstanding results on the well-known QALD-9, but did not generalize well on the recently appeared, structurally equivalent QALD-10~\cite{usbeck2024qald} and the BESTIARY benchmark (proposed by the authors) that obviously was not part of the model training.
Their results also showed that the model struggled to deal with an unknown \KG.
The inconsistent performance is explained by the possible memorization of the datasets by the models.

Some studies (\eg \cite{emonet2024llm,rangel2024sparql,lehmann2024large}) addressed an investigation of benefits from using \LLMs for domain-specific or \KG-specific tasks in \KGQA by leveraging, fine-tuning \LLMs, or prompting with intelligent few-shot selection.
Recent papers from Lehmann et al. \cite{lehmann2024beyond} and Zong et al. \cite{zong2024triad} leverage the novel LLM agent paradigm, where the model calls are streamlined into a predefined workflow and augmented with available tools -- external services that supply \LLMs with additional information.

From this brief analysis, we can conclude that, on one hand, combining \LLMs with \KGs might represent a promising technology for constructing \KGQA systems; on the other hand, well-known and often exploited benchmarks like \QALD, \LCQuAD provide the researchers with much better results.

\section{Approach and Materials}\label{sec:Mat_and_Met}
\newcommand{\highlight}[1]{\textbf{\emph{#1}}}
\subsection{Approach}\label{sec:approach}\label{sec:method}
Our primary research objective is to determine if state-of-the-art \LLMs are capable of generating valid \SPARQL queries over a \KG.
Our approach (presented in Fig.~\ref{fig:big-picture}) is tailored to answer the research questions while evaluating how well the \SPARQL query generation process performs with or without knowledge injection (\ie incorporation of external knowledge into models to improve their performance and/or influence their behavior).
To answer \ResearchQuestionA and \ResearchQuestionB, we carried out experiments for generating \SPARQL queries from \NL questions with \LLMs by using three prompting techniques.
For all \LLMs, we used the same prompts. 

The first type of prompt used in the experiments was \highlight{zero-shot prompting}.
This means that a question is sent to a model \qq{as-is} accompanied only by instructions to prevent extra text generation.
Fig.~\ref{fig:prompt1} illustrates the structure of this prompt. 


The second prompt, whose example is presented in Fig.~\ref{fig:prompt2}, was defined with the presumption that a Named Entity Recognition component was already executed and provided a perfect set of information about the named entities (including its linking to the \KG) within the given \NL question.
Here, the relevant entities were extracted from the gold standard queries and paired with the labels of each entity or property.
Pairs of the Entity Name and corresponding URI (in prefixed form) were injected into the prompt.
Therefore, we call this a \highlight{knowledge injection prompt}.

The third type of prompt (see Fig.~\ref{fig:prompt3}) was tailored to hide sensitive information from the model, trying to prevent the use of the memorized data.
Note, we additionally did not refer to the \Wikidata \KG in the prompts.
To mask the \Wikidata URIs, we changed them to random (yet unique) numbers with the SPARQL prefix \code{kg:} pointing to some graph that is unknown to the model.
Thus, in comparison to the knowledge injection prompt, all relevant information is recognizable from the prompt, however, there is no direct reference to \Wikidata.
Hence, we name this approach \highlight{masked knowledge injection prompt.}

We therefore infer that all three types of prompts supplement each other, so they are enough to answer the research question of this paper.
Using other types of prompts for investigating their influence of the \LLMs performance in generating \SPARQL queries from a \NL could be a subject of future research.

\begin{figure}[t!]
 
\begin{subfigure}[t]{0.99\textwidth}
    \captionsetup{singlelinecheck=false}
    \begin{lstlisting}
|\textcolor{blue}{Translate the question “Who developed Skype?” into a SPARQL query}|
|\textcolor{blue}{using the Wikidata Knowledge Graph.}|
|\textcolor{blue}{Have the query return only resources.}|
|\textcolor{blue}{Provide only the generated SPARQL query.}|
    \end{lstlisting}
    \vspace{-2ex}
    \caption{Example of zero-shot prompting.}
    \label{fig:llm-prompt}\label{fig:prompt1}
    \vspace*{2ex}
\end{subfigure}

\begin{subfigure}[t]{0.99\textwidth}
    \captionsetup{singlelinecheck=false}
    \begin{lstlisting}
|\textcolor{blue}{Translate the question "Who developed Skype?" into a SPARQL query}|
|\textcolor{blue}{using the Wikidata Knowledge Graph.}|
|\textcolor{blue}{Have the query return only resources.}|
|\textcolor{blue}{Provide only the generated SPARQL query.}|
|\textcolor{blue}{The possible entities and properties are:}|
|\textcolor{blue}{wd:Q40984 is Skype}|
|\textcolor{blue}{wdt:P178 is developer.}|
    \end{lstlisting}
    \vspace{-2ex}
    \caption{Example of a prompt with knowledge injection.}
    \label{fig:prompt2}
    \vspace*{2ex}
\end{subfigure}

\begin{subfigure}[t]{0.99\textwidth}
    \begin{lstlisting}[escapeinside=||,frame=single]{text}
|\textcolor{blue}{Translate the question "Who developed Skype?" into a SPARQL query}|
|\textcolor{blue}{using a Knowledge Graph.}|
|\textcolor{blue}{Have the query return only resources.}|
|\textcolor{blue}{Provide only the generated SPARQL query.}|
|\textcolor{blue}{The possible entities and properties are:}|
|\textcolor{blue}{kg:6211 is Skype}|
|\textcolor{blue}{kg:1548 is developer.}|
    \end{lstlisting}
    \vspace{-2ex}
    \caption{Example of a prompt with \qq{anonymized} knowledge injection. The information regarding linked URIs and masks was stored to be unmasked after a query generation.}
    \label{fig:prompt3}
\end{subfigure}
\caption{Examples of prompts to generate the SPARQL query 
for the question \qq{Who developed Skype?} from the \QALDNinePlus dataset by exploiting different LLMs.}
    \vspace*{-3ex}
\end{figure}

\subsection{Materials}
\subsubsection{Datasets}

During the past decade, researchers introduced to the community numerous \KGQA benchmarks with logical forms whose quantitative comparison was presented by Liu S.~et al.~\cite{liu-etal-2024-spinach} when introducing their own benchmark.

While executing our process, we aim also to find out how much the \LLM-generated output is affected by memorized data. 
Therefore, we suggest that the two benchmarks exploited are quite different in terms of using frequency and, respectively, in training datasets. 
We suppose that there is a higher tendency of \LLMs to memorize popular datasets and this should appear in the resulting metrics.
Hence, we require a popular dataset and a dataset that is rarely used where both are defined over data from the same \KG.

To evaluate the effectiveness of the proposed method, we carried out our experiments on two public benchmarks executable over the \Wikidata knowledge graph: \QALDNinePlus \cite{Perevalov22QALD9plus} and MCWQ \cite{cui2022compositional}.
Both datasets are multilingual; however, we used only their English parts for our current research.

\emph{\QALDNinePlus}\footnote{\url{https://github.com/KGQA/QALD_9_plus}}~\cite{Perevalov22QALD9plus} is an extension of the well-known \QALD-9 dataset where extended language support was added, and the translation quality for existing languages was significantly optimized (\eg via Spanish \cite{QALD-9-ES--A_Spanish_Dataset_for_Question_Answering_Systems-SEMANTiCS2023}) though the questions' translations were provided by multiple native speakers.
The dataset contains 558 questions incorporating information from the \Wikidata knowledge base.
In addition to the textual presentation, each question contains
the corresponding \SPARQL query, the answer entity URI, and the answer type.
The datasets of the \QALD series are widely used since more than a decade for scientific challenges and publications.

\emph{MCWQ}\footnote{\url{https://github.com/coastalcph/seq2sparql}}~\cite{cui2022compositional} is a rarely used dataset transformed from the CFQ (Compositional Freebase Questions) benchmark that was aimed at parsing questions into \SPARQL queries executable on the Freebase knowledge base.
The dataset has 2,444 question patterns (mod entities, verbs, etc.). 
There are 1,835 unique \SPARQL query patterns in MCWQ, resulting in 16.9\% instances covering 100\% of unique \SPARQL query patterns.
According to the authors, the dataset \qq{poses a greater challenge for compositional semantic parsing, and exhibits less redundancy in terms of duplicate patterns.}
In total, MCWQ contains 124,187 question query pairs, but our experiments were carried out on the gold test set. 
It should be pointed out that (1) the dataset has a focus on history and cinematography, (2) several questions express multiple restrictions to find the correct answer (sometimes much more restrictions than \QALDNinePlus has), and (3) multiple questions address the same topic but work with different restrictions.

\subsubsection{LLMs}
We employed 11 \LLMs from four developers in our experiments: Qwen 2.5~\cite{yang2024qwen2} (7B, 14B, 32B, and 72B), DeepSeek-r1~\cite{guo2025deepseek} (7B, 14B, 32B, and 70B), Mistral-Small and Mistral-Large\footnote{\url{https://mistral.ai/en/news/mistral-small-3}}, and \Llama 3.3 70B\footnote{\url{https://huggingface.co/meta-llama/Llama-3.3-70B-Instruct}} presented in Table~\ref{tab:LLMs_used}.
We chose these latest open-source models because of their significant variability in dimension, which allows us to evaluate different options on the selected benchmarks, as well as the high quality declared by developers.

The models were executed with the Ollama engine\footnote{\url{https://github.com/ollama/ollama}} on a server with two NVIDIA L40S GPUs (48GB VRAM).
All models are quantized in 4-bit with quantization type \qq{Q4\_K\_M} due to the limited resources.

\begin{table}[t]
\caption{Overview of the  LLMs used within our work}
\label{tab:LLMs_used}
\centering
\setlength{\tabcolsep}{4pt}
\scalebox{0.85}{
\begin{tabular}{l|c|c|c}
\toprule
    \textbf{Model name} & \textbf{Developed by} & \textbf{Year} & \textbf{Model size} \\
    \midrule
    Qwen 2.5 7B & Alibaba Cloud & 2024 & 7B \\
    Qwen 2.5 14B & Alibaba Cloud & 2024 & 14B \\
    Qwen 2.5 32B & Alibaba Cloud & 2024 & 32B \\
    Qwen 2.5 72B & Alibaba Cloud & 2024 &  72B \\
    DeepSeek-r1 7B & DeepSeek & 2025 & 7B \\
    DeepSeek-r1 14B & DeepSeek & 2025 & 14B \\
    DeepSeek-r1 32B & DeepSeek & 2025 & 32B \\
    DeepSeek-r1 70B & DeepSeek & 2025 & 72B \\
    Mistral-Small & Mistral AI & 2025 & 24B \\
    Mistral-Large & Mistral AI & 2025 & 123B \\
    \Llama 3.3 70B & Meta & 2024 & 70B\\
    \bottomrule
\end{tabular}
}
\vspace*{-2ex}
\end{table}

\subsubsection{Metrics}\label{sec:metrics}
For measuring the quality of the experimentally obtained results, we used two metrics: 1) the relative frequency of valid queries ($P_{val}$) and 2) precision ($P$) -- the relative frequency of correct answers.
Only a valid query can produce the correct answer, therefore, $P \le P_{val}$ applies. 
In our setting, a correct answer exists for all questions. 
This means that precision, recall, and \Fscore are always equal in one experiment. 


\section{Evaluation and Analysis}\label{sec:evaluation}
\subsection{Results}
Tab.~\ref{tab:results} reports the complete set of experiments on all models, covering all the metrics introduced in Sec.~\ref{sec:metrics}.
%
\newcommand*\rot{\rotatebox{90}}
\begin{table}[p]
\centering
\caption{Experimental results}
\label{tab:results}
\setlength{\tabcolsep}{4pt}
\resizebox{\columnwidth}{!}{%
\begin{tabular}{l||c|c|c|c|c||c|c|c|c|c||c|c|c|c|c}
\toprule
    \multirow{2}{*}{\textbf{Model}} & \multicolumn{5}{c||}{\textbf{Zero-shot}} & \multicolumn{5}{c||}{\textbf{Knowledge injection}} & \multicolumn{5}{c}{\textbf{Masked injection}} \\
    \cmidrule{2-16}
     & \rot{\textbf{Total records}} & \rot{\textbf{Valid queries}} & \rot{\textbf{Correct}} & \rot{\textbf{P\textsubscript{val}}} & \rot{\textbf{P = R = F1}} & \rot{\textbf{Total records}} & \rot{\textbf{Valid queries}} & \rot{\textbf{Correct}} & \rot{\textbf{P\textsubscript{val}}} & \rot{\textbf{P = R = F1}} & \rot{\textbf{Total records}} & \rot{\textbf{Valid queries}} & \rot{\textbf{Correct}} & \rot{\textbf{P\textsubscript{val}}} & \rot{\textbf{P = R = F1}} \\
    \midrule
    \multicolumn{13}{c}{\textbf{MCWQ}} \\
    \midrule

    Qwen 2.5 7B & 155 & 55 & 0 & 0.35 & 0.00 & 155 & 81 & 11 & 0.35 & 0.07 & 155 & 90 & 0 & 0.35 & 0.00 \\
    Qwen 2.5 14B & 155 & 45 & 0 & 0.29 & 0.00 & 155 & 66 & 6 & 0.29 & 0.04 & 155 & 78 & 0 & 0.29 & 0.00 \\
    Qwen 2.5 32B & 155 & 0 & 0 & 0.00 & 0.00 & 155 & 0 & 0 & 0.00 & 0.00 & 155 & 1 & 0 & 0.00 & 0.00 \\
    Qwen 2.5 72B & 155 & 18 & 0 & 0.12 & 0.00 & 155 & 95 & 24 & 0.12 & \textbf{0.15} & 155 & 65 & 2 & 0.12 & \textbf{0.01} \\
    DeepSeek-r1 7B & 155 & 2 & 0 & 0.01 & 0.00 & 155 & 11 & 0 & 0.01 & 0.00 & 155 & 4 & 0 & 0.01 & 0.00 \\
    DeepSeek-r1 14B & 155 & 76 & 0 & 0.49 & 0.00 & 155 & 26 & 2 & 0.49 & 0.01 & 155 & 45 & 0 & 0.49 & 0.00 \\
    DeepSeek-r1 32B & 155 & 59 & 0 & 0.38 & 0.00 & 155 & 10 & 6 & 0.38 & 0.04 & 155 & 9 & 0 & 0.38 & 0.00 \\
    DeepSeek-r1 70B & 155 & 69 & 0 & 0.45 & 0.00 & 155 & 12 & 2 & 0.45 & 0.01 & 155 & 49 & 2 & 0.45 & \textbf{0.01} \\
    Mistral-Small  & 155 & 92 & 0 & 0.59 & 0.00 & 155 & 71 & 11 & 0.59 & 0.07 & 155 & 38 & 0 & 0.59 & 0.00 \\
    Mistral-Large  & 155 & 135 & 1 & 0.87 & \textbf{0.01} & 155 & 112 & 12 & 0.87 & 0.08 & 155 & 110 & 0 & 0.87 & 0.00 \\
    \Llama 3.3 70B & 155 & 118 & 0 & 0.76 & 0.00 & 155 & 122 & 16 & 0.76 & 0.10 & 155 & 112 & 1 & 0.76 & \textbf{0.01} \\

    \midrule
    \multicolumn{13}{c}{\textbf{\QALDNinePlus}} \\
    \midrule
    Qwen 2.5 7B & 471 & 266 & 0 & 0.56 & 0.00 & 460 & 357 & 109 & 0.56 & 0.24 & 460 & 347 & 37 & 0.56 & 0.08 \\
    qwen2.5 14B & 471 & 268 & 1 & 0.57 & 0.00 & 460 & 342 & 209 & 0.57 & 0.45 & 460 & 246 & 126 & 0.57 & 0.27 \\
    Qwen 2.5 32B & 471 & 10 & 0 & 0.02 & 0.00 & 460 & 5 & 3 & 0.02 & 0.01 & 460 & 21 & 8 & 0.02 & 0.02 \\
    Qwen 2.5 72B & 471 & 247 & 6 & 0.52 & 0.01 & 460 & 400 & 257 & 0.52 & 0.56 & 460 & 425 & 229 & 0.52 & \textbf{0.50} \\
    DeepSeek-r1 7B & 471 & 31 & 0 & 0.07 & 0.00 & 460 & 84 & 3 & 0.07 & 0.01 & 460 & 30 & 0 & 0.07 & 0.00 \\
    DeepSeek-r1 14B & 471 & 293 & 0 & 0.62 & 0.00 & 460 & 198 & 77 & 0.62 & 0.17 & 460 & 46 & 4 & 0.62 & 0.01 \\
    DeepSeek-r1 32B & 471 & 354 & 4 & 0.75 & 0.01 & 460 & 347 & 212 & 0.75 & 0.46 & 460 & 335 & 181 & 0.75 & 0.39 \\
    DeepSeek-r1 70B & 471 & 370 & 3 & 0.79 & 0.01 & 460 & 144 & 64 & 0.79 & 0.14 & 460 & 55 & 14 & 0.79 & 0.03 \\
    Mistral-Small  & 471 & 338 & 6 & 0.72 & 0.01 & 460 & 399 & 241 & 0.72 & 0.52 & 460 & 347 & 151 & 0.72 & 0.33 \\
    Mistral-Large  & 471 & 447 & 30 & 0.95 & \textbf{0.06} & 460 & 450 & 279 & 0.95 & \textbf{0.61} & 460 & 426 & 212 & 0.95 & 0.46 \\
    \Llama 3.3 70B & 471 & 330 & 14 & 0.70 & 0.03 & 460 & 371 & 186 & 0.70 & 0.40 & 460 & 372 & 161 & 0.70 & 0.35 \\
    \bottomrule
\end{tabular}}
\end{table}
\begin{table}[p]
\centering
\caption{Experimental results for optimized MCWQ}
\label{tab:results_optimized}
\setlength{\tabcolsep}{4pt}
\resizebox{\columnwidth}{!}{%
\begin{tabular}{l||c|c|c|c|c||c|c|c|c|c||c|c|c|c|c}
\toprule
    \multirow{2}{*}{\textbf{Model}} & \multicolumn{5}{c||}{\textbf{Zero-shot}} & \multicolumn{5}{c||}{\textbf{Knowledge injection}} & \multicolumn{5}{c}{\textbf{Masked injection}} \\
    \cmidrule{2-16}
     & \rot{\textbf{Total records}} & \rot{\textbf{Valid queries}} & \rot{\textbf{Correct}} & \rot{\textbf{P\textsubscript{val}}} & \rot{\textbf{P = R = F1}} & \rot{\textbf{Total records}} & \rot{\textbf{Valid queries}} & \rot{\textbf{Correct}} & \rot{\textbf{P\textsubscript{val}}} & \rot{\textbf{P = R = F1}} & \rot{\textbf{Total records}} & \rot{\textbf{Valid queries}} & \rot{\textbf{Correct}} & \rot{\textbf{P\textsubscript{val}}} & \rot{\textbf{P = R = F1}} \\
    \midrule

    Qwen 2.5 7B & 146 & 51 & 0 & 0.35 & 0.00 & 140 & 77 & 12 & 0.35 & 0.09 & 140 & 70 & 0 & 0.35 & 0.00 \\
    Qwen 2.5 14B & 146 & 24 & 0 & 0.16 & 0.00 & 140 & 44 & 7 & 0.16 & 0.05 & 140 & 37 & 0 & 0.16 & 0.00 \\
    Qwen 2.5 32B & 146 & 0 & 0 & 0.00 & 0.00 & 140 & 0 & 0 & 0.00 & 0.00 & 140 & 1 & 0 & 0.00 & 0.00 \\
    Qwen 2.5 72B & 146 & 22 & 0 & 0.15 & 0.00 & 140 & 107 & 16 & 0.15 & \textbf{0.11} & 140 & 99 & 0 & 0.15 & 0.00 \\
    DeepSeek-r1 7B & 146 & 4 & 0 & 0.03 & 0.00 & 140 & 7 & 0 & 0.03 & 0.00 & 140 & 4 & 0 & 0.03 & 0.00 \\
    DeepSeek-r1 14B & 146 & 66 & 0 & 0.45 & 0.00 & 140 & 51 & 7 & 0.45 & 0.05 & 140 & 6 & 0 & 0.45 & 0.00 \\
    DeepSeek-r1 32B & 146 & 66 & 0 & 0.45 & 0.00 & 140 & 69 & 16 & 0.45 & \textbf{0.11} & 140 & 40 & 0 & 0.45 & 0.00 \\
    DeepSeek-r1 70B & 146 & 67 & 0 & 0.46 & 0.00 & 140 & 20 & 2 & 0.46 & 0.01 & 140 & 27 & 0 & 0.46 & 0.00 \\
    Mistral-Small  & 146 & 95 & 0 & 0.65 & 0.00 & 140 & 56 & 8 & 0.65 & 0.06 & 140 & 53 & 0 & 0.65 & 0.00 \\
    Mistral-Large  & 146 & 123 & 2 & 0.84 & 0.01 & 140 & 101 & 12 & 0.84 & 0.09 & 140 & 92 & 0 & 0.84 & 0.00 \\
    \Llama 3.3 70B & 146 & 103 & 0 & 0.71 & 0.00 & 140 & 97 & 20 & 0.71 & 0.14 & 140 & 89 & 0 & 0.71 & 0.00 \\

    \bottomrule
\end{tabular}}
\end{table}
For the sake of space, we will primarily focus on the top results.
Smaller models (7B and in some cases 14B) experienced difficulty in solving the zero-shot task and the masked injection task, especially regarding MCWQ.
The best overall quality demonstrated Qwen 2.5 72B, Mistral-Large, \Llama 3.3 70B, and DeepSeek-r1 32B, the worst one -- Qwen 2.5 32B.

Comparing on \Fscore in the experiments with knowledge injection prompting, both Mistral models (Mistral-Large$=0.61$ vs. Mistral-Small$=0.52$), and Qwen 2.5 72B ($0.52$), demonstrated the best results on \QALDNinePlus, while only the top result in the same experiment on MCQW only a $F1=0.15$ was achieved by Qwen 2.5 72B followed by \Llama 3.3 70B ($0.1$).
It is also noteworthy that even much smaller models of Qwen 2.5 demonstrated comparable results in the experiments where knowledge injection prompting was used.

However, the results of experiments with zero-shot prompting on both benchmarks are unsatisfactory: only Qwen 2.5 72B, Mistral-Large, Mistral-Small, DeepSeek-r1 32B, DeepSeek-r1 70B, and \Llama 3.3 70B were able to generate a few correct \SPARQL queries from the given \NL questions from the \QALDNinePlus dataset, while only Mistral-Large -- from MCWQ.

The experiment with masked knowledge injection prompting on MCWQ should be a challenge for all \LLMs. 
Only Qwen 2.5 72B, Mistral-Large, and \Llama 3.3 70B were able to generate one or two correct \SPARQL queries for 155 \NL questions.
Nevertheless, the same experiment on \QALDNinePlus unpredictably did not cause problems for most \LLMs besides Qwen 2.5 32B, DeepSeek-r1 7B, Deep\-Seek-r1 14B, and DeepSeek-r1 70B.

The general observation is that all metrics on the MCWQ benchmark are much worse when compared to \QALDNinePlus.
Trying to find out the reason for the poor quality of MCWQ results, we selectively analyzed questions from the dataset and ascertained that the fluency, semantic, and grammatical correctness of some questions do not conform to language norms.
Therefore, 140 questions were reformulated using OpenAI's well-performing GPT-4o\footnote{\url{https://platform.openai.com/docs/models\#gpt-4o}},  manually validated, and optimized the generated questions, \st a smaller, semantically equal, yet (language quality wised) improved dataset was generated\footnote{The dataset is stored in the online appendix including the generated \NL question and the manually optimized queries.} to evaluate the impact of the \textit{language quality of the \NL questions} in comparison to the memorization effect.
The results of the experiment with this improved MCWQ dataset are presented in Table~\ref{tab:results_optimized}.
However, they do not provide any noticeable improvement, and, therefore, refute the hypothesis that the MCWQ language quality has heavily impacted the LLM performance. 
Instead, we can conclude that the (missing) memorization effect was actually the reason for the observed SPARQL query generation quality.
It is worth noting that each request was carried out in a new conversation.


\subsection{Error Analysis}
We performed an analysis of erroneous queries by defining those not correspond to the expected response.
Research papers offer different error classifications (\eg \cite{lehmann2024large,diallo2024comprehensive,bhandiwad5009450bridging,lehmann2023language,taipalus2018errors}), however, we consider the error analysis as an additional result helping us to further evaluate the performance of our method.
We identified four error categories, which are presented with some examples in Fig.~\ref{fig:errors}.

\begin{enumerate}
    \item \textit{Invalid format or query} (\cf Fig.~\ref{fig:error-category-1}).
\begin{itemize}
\item 
    \textit{Invalid JSON format}.
    The model's output did not follow the correct JSON format, so the query text cannot be extracted.

\item 
    \textit{Invalid \SPARQL query}. 
    The query did not pass the syntax check (done with the help of RDFlib\footnote{\url{https://github.com/RDFLib/rdflib}}) or caused an error while being executed on \Wikidata.
\end{itemize}
\item \textit{Empty answer}  (\cf Fig.~\ref{fig:error-category-2}): all questions in the considered datasets would lead to a non-empty result from the Wikidata knowledge graph, hence at least one entity as an answer (or true/false value) is expected; therefore, an empty answer is an error.

\item \textit{Incorrect set of entities}  (\cf Fig.~\ref{fig:error-category-3}): we compared the set of expected entities with the set of produced entities.
In a correct answer, the query must produce the same list of entities as the gold standard, without regard to order.



\item \textit{Occurrence of \Wikidata URIs} (\cf Fig.~\ref{fig:error-category-4}): 
A \Wikidata URI occurred in the generated query, while the prompt requires URIs from the sample \KG (which is unknown to the \LLM).
\end{enumerate}

The error analysis clearly shows that all models are relying on the memorized data while generating \SPARQL query. 
In particular, a model was asked to generate a \SPARQL query for an imaginary \KG, but it produced the (correct) \Wikidata's URI instead (hence, Error Category 4 as presented on Fig.~\ref{fig:error-category-4}).

\begin{figure}[t!]
\centering

\begin{subfigure}[b]{0.99\textwidth}
\captionsetup{singlelinecheck=false}
\centering
\begin{lstlisting}
SELECT |{\color{dkblue}{?resource}}| 
WHERE { >// Instance of film
\end{lstlisting}
\vspace{-2ex}
\caption{Error Category 1 -- Invalid format/query}
\vspace{1ex}
\label{fig:error-category-1}
\end{subfigure}
\vfill

\begin{subfigure}[b]{0.99\textwidth}
\captionsetup{singlelinecheck=false}
\centering
\begin{lstlisting}
SELECT |{\color{dkblue}{?newSeriesEpisodes ?oldSeriesEpisodes}}| 
WHERE { 
    wd:Q162594 wdt:P1113 |{\color{dkblue}{?newSeriesEpisodes}}| .
    wd:Q180755 wdt:P1113 |{\color{dkblue}{?oldSeriesEpisodes}}| . 
ORDER BY DESC(|{\color{dkblue}{?newSeriesEpisodes}}|) 
LIMIT 1
}
\end{lstlisting}
\vspace{-2ex}
\caption{Error Category 2 -- Empty answer}
\vspace{1ex}
\label{fig:error-category-2}
\end{subfigure}
\vfill

\begin{subfigure}[b]{0.99\textwidth}
\captionsetup{singlelinecheck=false}
\centering
\begin{lstlisting}
SELECT |{\color{dkblue}{?mountain}}| 
WHERE  |{\color{dkblue}{?mountain}}| wdt:P31 wd:Q8502 .
SELECT (MAX(|{\color{dkblue}{?elevation}}|) AS |{\color{dkblue}{?maxElevation}}|) 
\end{lstlisting}
\vspace{-2ex}
\caption{Error Category 3 -- Incorrect set of entities}
\vspace{1ex}
\label{fig:error-category-3}
\end{subfigure}

\begin{subfigure}[b]{0.99\textwidth}
\captionsetup{singlelinecheck=false}
\centering
\begin{lstlisting}
SELECT |{\color{dkblue}{?moon}}|
WHERE { |{\color{dkblue}{?moon}}| kg:is_moon_of_Jupiter wd:Q61702557 .
|{\color{dkblue}{?moon}}| kg:is_mass wdt:P2067
}
\end{lstlisting}
\vspace{-2ex}
\caption{Error Category 4 – Occurrence of Wikidata URIs although a masked knowledge injection was used.}
\vspace{1ex}
\label{fig:error-category-4}
\end{subfigure}
\vspace*{-2ex}
\caption{Examples from the identified error categories in the generated SPARQL queries.}
\label{fig:errors}
    \vspace*{-3ex}
\end{figure}

Table \ref{tab:frequency} presents the frequency of erroneous queries for all \LLMs.
Investigating the data, we should point out that Error Category 1 (Invalid format or query) is more usually the case for the MCWQ dataset while Error Category 4 (occurrence of \Wikidata URIs) is rather relevant for \QALDNinePlus besides the DeepSeek-r1 models which demonstrate (excluding DeepSeek-r1 32B) high rate of Error Category 1 also when experimenting with \QALDNinePlus.
Error Category 3 (incorrect set of entities) is comparable for both datasets.

When analyzing the error categories by models, the noticeable fact is that Qwen 2.5 32B always produced invalid queries or did not follow the correct JSON format, \ie the query text cannot be extracted, while prompts to Mistral-Large hardly result in such errors. 
In addition, the DeepSeek-r1 models (besides DeepSeek-r1 32B) return the lowest number of empty answers (Error Category 2) on both datasets.
However, Mistral-Large often produces the incorrect set of entities, which is typically for all larger models (the size is 24B and over) on the \QALDNinePlus dataset and a small one -- Qwen 2.5 7B.
Therefore, the error frequency suggests that there is still much to be done to achieve acceptable results.

\newcolumntype{Y}{>{\centering\arraybackslash}X}

\begin{table}[t]
\caption{Categories of errors and their frequencies (note, multiple errors per query are possible).}
\label{tab:frequency}
\centering
\setlength{\tabcolsep}{4pt}
\scalebox{0.85}{
\begin{tabular}{|l||c|c|c|c||c|c|c|c|}
\hline
\multirow{3}{*}{Model} & \multicolumn{8}{c|}{Benchmarks/Error categories} \\ \cline{2-9}
& \multicolumn{4}{c||}{QALD-9-plus} & \multicolumn{4}{c|}{MCWQ} \\
\cline{2-9} 
& \multicolumn{1}{c|}{1} & \multicolumn{1}{c|}{2} & \multicolumn{1}{c|}{3} & \multicolumn{1}{c||}{4} & \multicolumn{1}{c|}{1} & \multicolumn{1}{c|}{2} & \multicolumn{1}{c|}{3} & \multicolumn{1}{c|}{4} \\
\toprule
Qwen 2.5 7B & 0.30 & 0.19 & \textbf{0.59} & 0.30 & 0.48 & 0.03 & 0.49 & 0.00 \\
Qwen 2.5 14B & 0.38 & 0.33 & 0.37 & 0.31 & 0.57 & 0.06 & 0.42 & 0.06 \\
Qwen 2.5 32B & \textbf{0.97} & 0.01 & 0.02 & 0.27 & \textbf{1.00} & 0.00 & 0.00 & 0.02 \\
Qwen 2.5 72B & 0.23 & \textbf{0.51} & 0.42 & \textbf{0.32} & 0.61 & 0.13 & 0.34 & 0.05 \\
DeepSeek-r1 7B & 0.90 & 0.05 & 0.10 & 0.16 & 0.96 & 0.01 & 0.04 & 0.02 \\
DeepSeek-r1 14B & 0.61 & 0.16 & 0.33 & 0.08 & 0.66 & 0.05 & 0.34 & 0.01 \\
DeepSeek-r1 32B & 0.26 & 0.45 & 0.46 & 0.29 & 0.81 & 0.02 & 0.18 & 0.00 \\
DeepSeek-r1 70B & 0.59 & 0.18 & 0.35 & 0.20 & 0.70 & 0.04 & 0.29 & \textbf{0.16} \\
Mistral-Small  & 0.22 & 0.43 & 0.49 & 0.29 & 0.55 & 0.06 & 0.43 & 0.00 \\
Mistral-Large  & 0.05 & 0.59 & 0.58 & \textbf{0.32} & 0.19 & \textbf{0.14} & \textbf{0.78} & 0.02 \\
\Llama 3.3 70B & 0.23 & 0.42 & 0.51 & \textbf{0.32} & 0.21 & 0.08 & 0.75 & 0.02 \\

\bottomrule
\end{tabular}
}
\vspace*{-2ex}
\end{table}

\section{Limitations and Discussion}\label{sec:limitations}
This paper focuses on the ability of state-of-the-art \LLMs to generate \SPARQL queries with or without knowledge injection.
However, when appraising the findings, it is essential to take into consideration some limitations and factors w.r.t. the general applicability of the proposed method.
First, we did not exploit commercial \LLMs like GPT, Claude, or other open-source \LLMs.
The exploited models are the newest ones, vary significantly in their dimensions, and demonstrate better results declared by developers. 
Second, we carried out our experiments only in English as the majority of research in this field is dominated by this language, and we paid most attention to \SPARQL queries here and not to the language capabilities.
On the other hand, the approach is portable to non-English languages, hence, we leave the exploration of this direction for future work.

A further limitation is that we conducted our experiments on two datasets.
Since the approach has been proven, the experiments could be repeated on other datasets, which might make a further contribution to the study of the memorization phenomenon.
However, it is worth noting that finding question-answering datasets that are rarely used is hard. 
Several questions are raised for the discussion.
First, synergistic integration of \LLMs with \KG (\ie \qq{knowledge injection} experiments) is still a promising direction for further research.
Even smaller models might provide benefits and, therefore, decent results to the users.
However, we have to point out the fact that all \LLMs demonstrated very poor quality when zero-shot-prompting.
Therefore, the exploited \LLMs are hardly able to generate \SPARQL queries from given \NL questions without knowledge injection.
So, one might come to the conclusion that the memorization effect needs to be triggered by a significant knowledge injection. 
Second, the results of all experiments on \QALDNinePlus outperform those on MCWQ.
All \LLMs struggled with this task on this infrequently exploited dataset, although the data provided via the knowledge injection prompting again provided all the needed information to generate a correct SPARQL query.
Our final experiment with an improved MCWQ dataset demonstrated that the linguistic quality of \NL questions does not impact the \LLM's performance significantly and, therefore, the final results.
In this regard, our results correlate with the ones from SPARQLGEN paper \cite{kovriguina2023sparqlgen}. 
Third, the error analysis revealed that during the \qq{masked injection} experiments, Wikidata's URIs occurred in the generated query while the prompt does not give any indicator that the generated query should be generated for the Wikidata \KG, instead it would have required URIs from the sample (unknown) \KG.
Hence, this evidence shows that memorized data significantly influences the output of \LLMs.
At the moment, there are still additional opportunities for process improvement.

\section{Conclusions and Future Work}\label{sec:conclusion}

In this study, we explored the capabilities of Large Language Models (LLMs) in generating SPARQL queries from natural language (NL) questions under three different conditions: zero-shot prompting, knowledge injection, and masked knowledge injection. 
Our primary goal was to assess the impact of structured knowledge integration on query generation performance, while also examining whether LLMs rely on memorization rather than true reasoning when dealing with knowledge graphs (KGs).

To address \ResearchQuestionA (\qq{Assuming a perfect knowledge injection, how well does the SPARQL query generation perform?}), our findings demonstrate that knowledge injection significantly enhances LLM performance. 
Models provided with explicit entity and property mappings consistently outperformed those operating under zero-shot conditions. 
This suggests that LLMs, despite their vast pre-training, still benefit greatly from additional structured context.
Even smaller models showed considerable improvement with knowledge injection, highlighting the importance of integrating external knowledge sources when leveraging LLMs for SPARQL query generation. 
However, performance varied among different models, with larger architectures generally yielding better results.

Regarding \ResearchQuestionB (\qq{What is the impact of memorization on the SPARQL query generation capabilities of LLMs}), our experiments revealed strong indications of memorization effects. 
When using masked knowledge injection, where original Wikidata entity URIs were anonymized to prevent recognition, several models still generated queries containing correct Wikidata URIs. 
This suggests that LLMs often rely on memorized training data rather than true reasoning capabilities, raising concerns about their generalizability to unseen datasets or novel KGs. 
Moreover, models performed significantly better on the well-known \QALDNinePlus dataset compared to the less frequently used MCWQ dataset, reinforcing the notion that model familiarity with a dataset influences results.

In conclusion, researchers and practitioners need to be cautious when using \LLM-based SPARQL generation approaches and always keep in mind that the results of the presented methods are unlikely to be fully reproducible on a private or new dataset because the memorization effect will significantly increase the quality and consequently degrade the results in a dataset without this effect.
Further research will be necessary to compensate for this effect.

\textbf{Future Work.} 
While our study provides valuable insights into SPARQL query generation with LLMs, several limitations should be considered. 
First, our experiments were conducted using a large but still selected set of open-source models and two benchmark datasets. 
Further research could expand the scope to involve additional models, including proprietary LLMs such as GPT, Claude, or Gemini, to determine whether similar trends persist. 
Second, our experiments were performed exclusively on English questions. 
Given that many knowledge graphs contain multilingual data, assessing LLM performance across different languages remains an important avenue for future exploration.
Third, the detailed error analysis and an entire error classification could provide the researchers with ways to improve the performance of \LLM-based \SPARQL generation approaches.

Additionally, our findings highlight the need for improved strategies to mitigate memorization effects and the new strategies for creating and publishing benchmarking datasets. 
Future research should investigate fine-tuning methods or hybrid approaches that combine LLMs with rule-based or symbolic reasoning techniques to enhance generalization. 
Another promising direction is the integration of external knowledge retrieval mechanisms, such as KG-based embeddings or reinforcement learning strategies, to improve query accuracy while minimizing reliance on pre-trained data.

Finally, evaluating LLM performance across a broader range of KGQA tasks (\eg handling complex queries, reasoning over multiple hops, and supporting federated queries) could provide further insights into their real-world applicability. 
By addressing these challenges, we can advance the development of more reliable, interpretable, and generalizable LLM-driven approaches for knowledge graph-based question answering.

\newpage
\bibliographystyle{splncs04}
\bibliography{bibl}
\end{document}